\documentstyle[12pt]{article}
\begin{document}

\title{
\Large \bf  Decoding Information by Following  Parameter Modulation
With Parameter Adaptive Control}

\author{ Changsong Zhou$^1$ and  C.-H. Lai$^{1,2}$ \\
        $^1$Department of Computational Science\\
        and $^2$Department of Physics\\
        National University of Singapore,
        Singapore 119260}

\date{}
\maketitle
\baselineskip 12pt
\vspace{0.5cm}

\begin{center}
\begin{minipage}{13cm}
\baselineskip 12pt
\centerline{\bf Abstract}

It has been proposed  to realize secure communication using chaotic
synchronization via transmission of binary message encoded by parameter
modulation in the chaotic system.   
This paper considers the use of parameter adaptive control techniques to
extract the  message, based on the assumptions that we know the equation
form of the chaotic system in the transmitter but do not have access to the
precise values of the parameters which are kept secret as a secure set. 
In the case that a synchronizing system can be constructed using  parameter 
adaptive control by the transmitted signal and the synchronization is robust to
parameter mismatches, the parameter modulation can be revealed and the message
decoded without resorting to exact parameter values  in the secure set.
A practical local Lyapunov function method for designing parameter
adaptive control rules based on originally synchronized systems is presented.


\end{minipage}
\end{center}
\newpage
\vspace{1cm}

\section{Introduction}
\bigskip
Chaotic dynamics, which have noise-like  broadband power spectra,
are interesting candidates for encoding and masking information signals
in communication. Most approaches proposed to realize  this basic idea of
using a chaotic signal as broadband carrier are  based  on the
synchronization of coupled chaotic systems~\cite{pc}.  Several 
schemes have been 
proposed so for: (I) {\sl  chaotic masking}~\cite{khec,co} where the message is added directly 
to the chaotic carrier with an amplitude much lower than the chaotic carrier;
(II) {\sl chaotic modulation}~\cite{kp,pksp,kps}  where the message is injected 
into the chaotic system to modulate the chaotic carrier, and (III)
{\sl chaotic switching}~\cite{co,pckh,dkh} where a binary 
message is transmitted by switching between two chaotic attractors associated
with two sets of  parameters of the system.  

Intuitively, the communication is expected to be secure based on two
considerations: (1) it is difficult to
read out the hidden message by any spectral analysis due to the broadband
nature of the chaotic carrier, and (2) exact knowledge of the parameters of the
system in the receiver is necessary to recover the message. 
Thus, a set of the system parameters which serve as the encryption key, is not
accessible by any third party. 
 
However, recently some researchers  have shown that the security
may be spoiled, not by access to the secure set of the parameters, but by some
other approaches. For the communication schemes (I) and (II),  
it has been shown that the hidden 
message may be unmasked with some nonlinear dynamical forecasting 
methods~\cite{short1,short2}. 
It is  believed that  this weakness in security is due to low 
dimension and single positive Lyapunov exponent of the chaotic carrier, 
and the suggestion is  to employ hyperchaotic
systems, such as coupled chaotic arrays~\cite{kps,pdy} or time-delay
 systems~\cite{ml,glp} in communication. 
This however may not produce drastic improvement in the security, as 
shown in a recent report~\cite{short3} that messages masked  by hyperchaos 
 of a six-dimensional  system can be unmasked only with a three-dimensional 
reconstruction,  and in our recent work~\cite{zl} demonstrating that
 messages masked 
by chaos of time-delay systems with very hige dimension and many 
positive Lyapunov exponents can also be extracted successfully. 

Another work~\cite{pc1} has also  shown that hidden messages  can  be extracted
from  chaotic carrier without  reconstructing
the full dynamics, but using  some suitable return maps, which is successfully applied to
the Lorenz system for communication schemes I and III. 

The idea of encoding by parameter modulation is to use 
two chaotic attractors ${\cal{A}}$ and ${\cal{B}}$ to 
represent the two symbols of the digital signals~\cite{co,pckh,dkh}.  
Since ${\cal{A}}\not={\cal{B}}$, it is possible to construct some suitable
return maps which can distinguish the differences between the
two attractors, thus reading  out the message, just as shown in 
Ref.~\cite{pc1}.
However, if the two attractors are rather complex or the differences
between them are very subtle, it may  be
very difficult  to find such distinguishable return maps.  

It is natural to ask if it is possible for a motivated intruder to follow 
 the parameter modulation in the 
transmitter using some parameter adaptive control by the transmitted signal.
This paper  carrys out security analysis of communication
systems using the encoding scheme III. Our analysis is based on the following assumptions: 
\begin{description}
\item (a) The intruder does not have access to the precise value of any system
           parameters in the secure set.  
\item (b) The intruder does know the functional form of the chaotic system in the
            transmitter.
\end{description}

Our results will show that if a synchronizing system can be 
constructed using parameter
adaptive control by the transmitted signal and
 the synchronization is robust to parameter 
mismatches, the messages may be decoded without resorting to the 
exact parameter values.
Since it is practically possible for a motivated intruder to locate a region close to 
the exact parameters based on the knowledge of the system and the character of the
transmitted signal, the security may be spoiled. Robustness of the 
synchronization to parameter mismatches is an advantage for implementation of the 
communication scheme but may be a weakness to the security.

\bigskip

\section{Decoding by parameter adaptive control}

Let us consider the following transmitter system
\begin{equation}
\frac{d{\bf x}}{dt}={\bf F}({\bf p, q, x}),
\end{equation}
where ${\bf p}$  and ${\bf q}$ are system parameters. Binary messages are 
encoded by switching ${\bf q}$ between ${\bf q}_1$ and ${\bf q}_2$.   
Based on our assumption, a motivated intruder has known the equations of the system
and that ${\bf q}$ are modulated for encoding. He tries  to construct a decoding system 
using parameter adaptive control by the transmitted signal $s=h({\bf x})$, as 
\begin{eqnarray}
\frac{d{\bf y}}{dt}&=&{\bf H}({\bf p}_y, {\bf q}_y, {\bf y}, s),\\
\frac{d{\bf q}_y}{dt}&=&{\bf G}({\bf y}, {\bf q}_y, s)(s-h({\bf y})),
\end{eqnarray}
Suppose that this system (called as the {\sl intruder} system from now on) is
synchronizable with some suitable coupling function ${\bf G}$ if ${\bf p}_y={\bf p}$. 

In general, it is not always possible that one can find a synchronizable intruder system 
for any transmitted signal $s$ and any subset ${\bf q}$ of the system parameters. 
However, if with 
this transmitted signal $s$, a synchronizable system ${\bf H}({\bf p}, {\bf q}, {\bf y}, s)$ 
 can be found by some synchronization schemes,
such as Pecora-Carroll decomposition~\cite{pc}, active-passive decomposition~\cite{kp} 
or feedback control~\cite{pyr}  {\sl without} parameter 
adaptive control, then we can expect that additional parameter adaptive control loops for 
parameters ${\bf q}_y$ {\sl exist} if the synchronization is robust to parameter 
mismatches to some extent, because the system ${\bf H}$ driven by $s$ is still stable  
(the largest conditional Lyapunov exponent is negative) for parameters ${\bf q}_y$
 in the vicinity of the point 
${\bf q}_y={\bf q}$ although exact synchronization is spoiled by parameter mismatches, and 
the exact synchronization can be restored by bringing the parameters back to the point ${\bf q}_y={\bf q}$
using some appropriate control methods.  For some systems, such  parameter control rules can be
found by an  analysis based on a global
Lyapunov function. In general however, such an analytical treatment may not be possible. In this case,
we employ the idea of designing control 
rule using the information of a control surface~\cite{ps} constructed by  perturbing the parameters ${\bf q}_y$.  
The essence of the idea  is that since the synchronization between the systems  ${\bf H}$ and ${\bf F}$
 before incorporating 
parameter adaptive control is robust to parameter mismatches, the synchronization behavior changes smoothly 
when ${\bf q}_y$ deviate slightly from ${\bf q}$. And there exists  a local Lyapunove  function   
 with respect to  the parameters ${\bf q}_y$ near ${\bf q}$, with the form   
\begin{equation}
E({\bf q}_y)= {\bf U}^T{\bf U}      
\end{equation}
where ${\bf U}=(U_1, U_2,\cdots, U_k)^T$ are time average of some functions
${\bf u}=(u_1, u_2,\cdots, u_k)^T$ ($k$ is the number of the components in  ${\bf q}_y$), i.e., 
\begin{equation}
U_i=\lim\limits_{T\to \infty}\frac{1}{T}\int_0^T u_i dt \;\;\;(i=1,2,\cdots, k).
\end{equation}
The  function $u_i$  has  the form 
$u_i=\hat{u}_i(s,{\bf y},{\bf q}_y)(s-h({\bf y}))$ so that ${\bf U}={\bf 0}$ if ${\bf q}_y={\bf q}$. 
In order that $E({\bf q}_y)$
is a local Lyapunov function, it is required that ${\bf U}$ is smooth with respect to ${\bf q}_y$ near ${\bf
q}$. With this local Lyapunov function, the following  evolution system
\begin{equation}
\frac{d{\bf U}}{dt}=-\alpha {\bf U},
\end{equation} 
is stable at ${\bf U}={\bf 0}$. $\alpha$ is a convergence parameter.
Noting that the convergence of ${\bf U}$ is induced by control of the parameters, we have 
\begin{equation}
\frac{d{\bf q}_y}{dt}=\frac{\partial {\bf q}_y}{\partial {\bf U}}\frac{d{\bf U}}{dt}= 
-\alpha \frac{\partial {\bf q}_y}{\partial {\bf U}}{\bf U}.
\end{equation}
In general,  it is impossible to obtain an explicit form of ${\bf U}$ for the control rule in Eq. (7). 
To solve this problem, we use a control surface obtained in simulation or experiment. 
First  we  record a time series $s$ from system ${\bf F}({\bf p, q, x})$ with a known set of parameter 
$({\bf p, q})$ in the chaotic regime. Then we perturb the parameter ${\bf q}_y$ in the driven system
${\bf H}({\bf p}, {\bf q}_y, {\bf y}, s)$ slightly from ${\bf q}$ to some 
values in its vicinity, and compute ${\bf U}({\bf q}_y)$ as a function of ${\bf q}_y$. For appropriately chosen 
function ${\bf u}$, ${\bf U}({\bf q}_y)$ are smooth  
with respect to  ${\bf q}_y$  close to the point ${\bf q}$, and in the  vivinity, 
 it  can be approximated by 
\begin{equation}
{\bf U}({\bf q}_y)=M({\bf q}_y-{\bf q}),
\end{equation}
where the constant $k\times k$ matrix $M$ is  obtained by a local linear fitting of the 
numerically or experimentally obtained control surface ${\bf U}({\bf q}_y)$.  Now if
 the initial value of ${\bf q}_y$ is close to ${\bf q}$, we can replace
the Jacobian matrix $\partial {\bf q}_y/\partial {\bf U}$ in Eq. (7) by 
the matrix $M^{-1}$, i.e.
\begin{equation}
\frac{d{\bf q}_y}{dt}=-\alpha M^{-1}{\bf U}.
\end{equation}
In practice, one can implement the above control rule by replacing ${\bf U}$ with a time average over
a period of time $\tau$, e.g.
\begin{equation}
U_i(t)=\frac{1}{\tau }\int^t_{t-\tau} u_i dt.
\end{equation}
Often, the parameter modulation in the transmitter is much slower than the time scale of the 
chaotic system, and 
one can simplify the control rule by replacing the time average $U_i$ with $u_i$, and finally obtains
\begin{equation}
\frac{d{\bf q}_y}{dt}=-\alpha M^{-1}{\bf u}.
\end{equation}

We can expect that  with the above additional parameter adaptive control, the 
synchronization between the systems is maintained with small enough coupling strength
$\alpha$ for ${\bf q}_y$ initially close to ${\bf q}$, so that  the parameters ${\bf q}$ can be recoved. 
We can also expect that the synchronization is also robust to mismatches of the  rest  
parameters ${\bf p}$. The function $u_i$ can be chosen somewhat arbitrarily, as long as $U_i$ 
(and thus $E({\bf q}_y)$) is a smooth function of ${\bf q}_y$ in the neighborhood of ${\bf q}$. 
This scheme provides a  general and  practical yet  simple way to build  additional
 parameter adaptive control loops for  originally coupled and synchronized systems, 
even though a proper choice of
the functions $u_i$ may still be  nontrivial.   
In this way, the intruder can design systematically an attacking system for the communication scheme based on
the knowledge of the system, 
although so designed control scheme using local information  may not be successful 
when applied to the signal from the transmitter
whose parameters the intruder does not know, especially when the transmitter is operating in a parameter region
far way from that  the intruder uses to build the control rule.  However, it is possible for the intruder
to get into a  neighborhood of the transmitter parameters using some system identification methods
based on the knowledge of the system.    

Since  the intruder system  is 
quite robust to parameter mismatches, the parameter modulation in the transmitter  may be
revealed  and the message decoded  without resorting to the exact values of the transmitter parameters ${\bf p}$, 
but within some tolerable neighborhood.  
In certain cases, it might also be possible
to recover all the transmitter parameters $({\bf p, q})$ 
by designing adaptive control loops for all of  them with the above 
scheme.    

In the next sections, we present examples of message decoding  based on 
the above idea. In the first example of Chua's circuit, the control rule is  obtained
with a global function analysis and in the second example of Lorenz system, the control rule 
is designed with the local Lyapunov function scheme.  

\section{An example of Chua's circuit}

As an illustration, we carry out analysis on a specific communication system 
proposed in Ref.~\cite{dkh}. We first give  a brief description of 
the encoding scheme, and then construct a robust intruder system. 

\subsection{The transmitter}

The communication system employs the well-known Chua's circuit as
the chaotic system. 
The evolution equations for the  Chua's circuit are given by
\begin{eqnarray}
C_1\frac{dx_1}{dt}&=&\frac{1}{R}(x_2-x_1)-h(x_1),\\
C_2\frac{dx_2}{dt}&=&\frac{1}{R}(x_1-x_2)+x_3,\\
L\frac{dx_3}{dt}&=&-x_2.
\end{eqnarray}
The nonlinear characteristic of the Chua's diode $h(x_1)$ is given by
\begin{equation}
h(x_1)=G_1x_1+\frac{1}{2}(G_0-G_1)[|x_1+B_p|-|x_1-B_p|],
\end{equation}
which is a three-segment piecewise-linear function.

A binary message stream $I_{in}$ is encoded by switching between parameters  
$G_0$ , $G_1$ and $G_0^{\prime}=G_0+\frac{1}{r}$ , $G_1^{\prime}=G_1+\frac{1}{r}$ when
the stream  switches between  $+1$ and $-1$,  where $r$ is a
resistor  in parallel with Chua's diode. The parameters used are shown in Table I. 
Since  $1/r$ is small (about $1\%$ with respect to $G_0$ and $G_1$), 
the two chaotic attractors are very similar, as shown in
Fig. 1. To examine the similarity, we construct return maps  using the 
consecutive maxima  $x_{max}(n)$ and $x_{max}(n+1)$ from the transmitter 
signal $x_1$, as done in Ref.~\cite{pc1}.   
The results are shown in Fig. 2, with circles for $I_{in}=1$ and crosses for
$I_{in}=-1$ respectively. It is seen that the maps are quite complicated, and 
most of the points of the two maps coincide and entangle with each other.
Distinguishable  difference between the two maps is only seen for $x_{max}(n)$
around $-0.5$, which consists of only a small fraction of the maxima. Extracting
the message, although is not totally impossible, can be done only for  
 a small portion of the message bits.

\bigskip 
\subsection{The intruder system}

Based on our assumption, the intruder does know that the chaotic system of the
transmitter is the Chua's circuit and  that the message is encoded by the modulation of 
$G_0$ and $G_1$, 
but does not know the value of any of the
parameters $C_1, C_2, R, L, B_p, G_0, G_1, r$.
Based on the available information, the intruder constructs a receiver system  
based on the idea of parameter adaptive control as follows:  
\begin{eqnarray}
C_1\frac{dy_1}{dt}&=&\frac{1}{R}(y_2-y_1)-h(x_1),\\
C_2\frac{dy_2}{dt}&=&\frac{1}{R}(y_1-y_2)+y_3,\\
L\frac{dy_3}{dt}&=&-y_2,\\
\frac{dQ_0}{dt}&=&\frac{1}{2}x_1[y_1-x_1][1-{\hbox{sgn}}(|x_1|-B_p)],\\
\frac{dQ_1}{dt}&=&\frac{1}{2}x_1[y_1-x_1][1+{\hbox{sgn}}(|x_1|-B_p)].
\end{eqnarray}
where $Q_0$ and $Q_1$ are controllable  parameters of $h(x_1)$ which is
now
\begin{equation}
h(x_1)=Q_1x_1+\frac{1}{2}(Q_0-Q_1)[|x_1+B_p|-|x_1-B_p|].
\end{equation}
Eqs. (19-20) mean that $Q_0$ is  modified when $|x_1|\leq B_p$ and
$Q_1$ is modified when $|x_1|> B_p$.

The intruder system of Eqs.~(16-21) will synchronize with  the transmitter 
Eqs.~(12-15) if the 
 parameters $C_1, C_2, R, L, B_p$ are identical for the two systems.
To prove it, let us examine the dynamics of the difference
$e_i=y_i-x_i (i=1,2,3)$, $e_4=Q_0-G_0$ and $e_5=Q_1-G_1$ given by
\begin{eqnarray}
C_1\frac{de_1}{dt}&=&\frac{1}{R}(e_2-e_1)
-\frac{1}{2}x_1e_4[1-{\hbox{sgn}}(|x_1|-B_p)]
-\frac{1}{2}x_1e_5[1+{\hbox{sgn}}(|x_1|-B_p)],\\
C_2\frac{de_2}{dt}&=&\frac{1}{R}(e_1-e_2)+e_3,\\
L\frac{de_3}{dt}&=&-e_2,\\
\frac{de_4}{dt}&=&\frac{1}{2}x_1e_1[1-{\hbox{sgn}}(|x_1|-B_p)],\\
\frac{de_5}{dt}&=&\frac{1}{2}x_1e_1[1+{\hbox{sgn}}(|x_1|-B_p)].
\end{eqnarray}
The global Lyapunov function
\begin{equation}
E=C_1e_1^2+C_2e_2^2+Le_3^2+e_4^2+e_5^2,
\end{equation}
\begin{equation}
\frac{dE}{dt}=-\frac{2}{R}(e_1^2+e_2^2)\leq 0,
\end{equation}
suggests that  the state and parameters of the intruder system  will
converge to those of the transmitter. Fig. 3 illustrates  the synchronization
process of the system with  the attractor in Fig. 1(a) ($I_{in}=1)$. 
The synchronization error decreases exponentially with time, with fluctuations 
only within small time scales, and  we can expect that the synchronization is robust to 
parameter mismatches. 
Note that the  stable values $Q_0=-0.753$ and $Q_1=-0.396$ are just
the values of $G_0$ and $G_1$ in the  transmitter.

When the information stream enters the transmitter, lasting a time interval
$T$ for each bit, the transmitted
signal is a sequence switching between  the two chaotic attractors in Fig. 1.
We take  $T=4.65$ ms  as in Ref.~\cite{dkh}. With the transmitted signal $s=x_1$ 
(Fig. 4(a)) carrying   a  random message stream, the intruder observed the change of 
parameter $Q_0$ and $Q_1$ in Fig. 4(b) and (c) respectively. 
Switching between the two chaotic attractors results in only small
fluctuations of $Q_0$, but sudden jumps of $Q_1$, because $|x_1|>B_p$ most of the
time so that $Q_1$ is modified more frequently. 
After  a transient
of about $50$ ms,  $Q_1$ comes  to oscillate slightly  about $-0.395$ for bit
$+1$ and $-0.385$ for bit $-1$.  
 A comparison between the  evolution of $Q_1$ and the parameter modulation 
in the transmitter shows clearly
that the message can be decoded correctly except for a few bits during the 
synchronization transient. 
 An interesting thing  is that, since $T$ is much 
smaller than the  relaxation time of synchronization (about 50 ms, see Fig. 3), the 
intruder operates in a regime of synchronization transient after the message stream
switches from one value to the other.  As a result, the oscillation amplitude 
of $Q_1$ (about 10 $\mu$s) is larger than the parameter 
modulation $\frac{1}{r}=6 {\mu}$s in the transmitter, which can be an advantage for 
message decoding.   

So far, we use the exact values of the transmitter parameters
$C_1, C_2, R, L, B_p$ in the intruder system  to demonstrate that it  is 
able to follow the parameter modulation in the transmitter by adaptive control. 
By assumption, the intruder does not have access to these values. However, it is
possible to locate an approximate region in the parameter space using some 
characteristic quantities for system identification based on the knowledge of the 
chaotic system and  at the same time monitor the synchronization error during the scanning
of the parameter space.    
In the following simulation, we suppose that the intruder is able to locate a region
within $20\%$ deviation from the precise values of the parameters.  
 We choose 5 random values in $[-0.2, 0.2]$ as the
relative difference of the above parameters between the intruder  and the transmitter,
as displayed in table II as an example. 
For the same information stream in Fig. 4(c), the evolution of $Q_1$ is now
presented in Fig. 5(a) which is  shifted to oscillate quite noisely  around $-0.36$ 
 due to the parameter mismatches. 
After smoothing the fluctuations with a moving
average filter with length of 4 ms,  the oscillation of $Q_1$ reveals most of 
the parameter modulation in the transmitter correctly, as seen in Fig. 5(b). 
We use a simple threshold
testing to recover the message, as shown in Fig. 5(c) with a threshold 
$Q_{th}=-0.365$. 
A comparison between the recovered and the original messages has clearly shown that 
the security is compromised.  
The results are also robust to external noise, as seen in Fig. 6 for the 
same parameters
as in Fig. 5, but with $x_1$ containing noise between $[-0.2,0.2]$.

\section{ An example of Lorenz system}

In the above example, we are able to write down the parameter adaptive control rules based
on an analytical treatment by a global Lyapunov function. In the following example, we revisit 
the communication system in Refs.~\cite{co, pc1} to illustrate the idea of designing an intruder system using 
the above local Lyapunov function method, although it has been shown that the message can be extracted
using some  suitable return maps~\cite{pc1}.

The communication system using Lorenz system is
\begin{eqnarray}
\frac{dx_1}{d\tau}&=&\sigma(x_2-x_1),\\
\frac{dx_2}{d\tau}&=&rx_1-x_2-x_1x_3,\\
\frac{dx_3}{d\tau}&=&x_1x_2-bx_3,
\end{eqnarray}
where  $\sigma=16.0, r=45.6$ and  $b$ is modulated between $b=4.0$ and $b=4.4$. $s=x_1$ is the transmitted signal.

We can design an attacking  system with parameter adaptive control for parameter $b$
 based on the following system
coupled by feedback~\cite{pyr}, 
\begin{eqnarray}
\frac{dy_1}{d\tau}&=&\sigma(y_2-y_1)+\epsilon (s-y_1),\\
\frac{dy_2}{d\tau}&=&ry_1-y_2-y_1y_3,\\
\frac{dy_3}{d\tau}&=&y_1y_2-by_3,
\end{eqnarray}
which will be synchronized  with the system $x$ for large enough coupling strength $\epsilon$. 
The synchronization is also quite robust to parameter 
mismatches for large $\epsilon$. Since by assumption,  we  do not know the 
parameter values in the transmitter,  
we uses $(\sigma, r, b)=(10, 28, 8/3)$ from a chaotic region in experiment or simulation. With $\epsilon=40$,
for example, the two system is synchronized. Now let systems $x$ and $y$ have the same  $\sigma$ and 
$r$, but perturb the parameter  $b$ in the system $y$ around $b=8/3$, e.g. $b_y=b(1+\Delta)$.
We calculate $U(\Delta)$  as a function of $\Delta$ by trying the following  three simplest functions
$u_1=(s-y_1)y_1$, $u_2=(s-y_1)y_2$ and $u_3=(s-y_1)y_3$ . The results of $U$ are shown in Fig. 7. 
It is seen that $U$ is  a smooth function    
for $\Delta$ close to zero for  the  functions  $u_1$ and $u_2$, but not for $u_3$. 
And it is obvious that $U$ is also a smooth function for any linear
combination of $u_1$ and $u_2$.  Some other choices of $u$ is possible, for example $u=(s-y_1)y_1y_3$. 
Now we can introduce  the additional parameter control loop
\begin{equation}
\frac{db}{d\tau}=\alpha u,
\end{equation}
where $u$ can be $u_1$, $u_2$ or their any linear combinations or many other possible choices. 
The sign of $\alpha$ is determined by the sign of $dU/d\Delta$. 
Simulations have  demonstrated that so designed control rules maintain the synchronization for $\alpha$ small 
enough. $\alpha$ is allowed to be larger for larger $\epsilon$. The control is still stable if the
system parameters are shifted from $(\sigma, r, b)=(10, 28, 8/3)$ to those of the transmitter, and the initial
values of the parameter $b$ in the system $y$ does not need to be close to that of the system $x$.  
Since our
purpose is to illustrate the designing idea, we do not go into great details on the synchronization behavior 
in the parameter space $(\epsilon, \alpha)$. The fact is, in a large region of this parameter space, the 
synchronization is very robust to mismatches of the rest parameters $\sigma$ and $r$.    
An example of the synchronization without and with the additional parameter control loop is shown in Fig. 8 
for $u=u_1=(s-y_1)y_1$, $(\sigma, r, b)=(16.0, 45.6, 4.0)$ and $(\epsilon, \alpha)=(40,0.1)$. The synchronization is a little 
slower when parameter adaptive control is introduced, and it is robust to parameter mismatches because 
the synchronization error decreases exponentially, with fluctuations only within very small time scales. 
The parameter $b$ comes very close to $b=4.0$ in the transmitter  from a large initial value $b=17.0$ within only a few ms (in the
new time scale below). 

Now let us use the  system to attack the secure communication.
 In order that  the time scale agrees with the system in~\cite{co,pc1},
we introduce a new time scale $t=\tau/K$ where $K=2505$~\cite{pc1} is a scale factor. In the transmitter, the bit duration 
is $4$ ms. As in the above section, we first demonstrate the parameter recovery for identical parameters
$\sigma$ and $r$ in the transmitter and intruder systems, as seen in Fig. 9.  
Then, we examine the robustness to the parameter 
mismatches of $\sigma$ and $r$. The message can be recoved quite reliablely if $(\sigma_y, r_y)$ in the intruder 
system is within a relatively close neighborhood of the transmitter, say, within a $20\%$ deviation. Message 
decoding is generally  extremely robust for  $\sigma_y<\sigma$ and $r_y>r$. 
An example for $\sigma_y=0.37\sigma$ and
$r_y=1.72r$ is shown in Fig. 10. It is seen that $b$ has been made to oscillate around $b=2.2$ rather
than $b=4.2$ in the transmitter due to very large parameter mismatches, however, the message is recoved
correctly with a moving average filter with length of 2 ms and  a simple threshold test.      

In the following, we go further to  design adaptive control  
loops for all the three parameters $(\sigma, r, b)$ in the Lorenz system.
We find that ${\bf U}$ changes smoothly when the parameters in the originally synchronized systems 
change slightly if we choose  the following three functions 
\begin{equation}
u_1=(s-y_1)y_1y_3, \;\;\; u_2=(s-y_1)y_2, \;\;\; u_3=(s-y_1)(y_1+y_2)
\end{equation}
The control surface is obtained by perturbing the parameters  in the driven system  within a $2\%$
vicinity of $(\sigma, r, b)=(10, 28, 8/3)$. $U_i (i=1,2,3)$ is the time average of $u_i$ in a period
of 0.1 second in the time scale $t$.  
After evaluating the matrix $M^{-1}$, we obtained the following attacking system with 
parameter adaptive control loops:
\begin{eqnarray}
\frac{dy_1}{d\tau}&=&\sigma(y_2-y_1)+\epsilon (s-y_1),\\
\frac{dy_2}{d\tau}&=&ry_1-y_2-y_1y_3,\\
\frac{dy_3}{d\tau}&=&y_1y_2-by_3,\\
\frac{d\sigma}{d\tau}&=&\alpha(-0.293u_1-18.5u_2+15.7u_3),\\
\frac{dr}{d\tau}&=&\alpha(1.18u_1+95.4u_2-75.4u_3),\\
\frac{db}{d\tau}&=&\alpha(-0.123u_1-10.2u_2+8.10u_3).
\end{eqnarray}
This system is synchronized  for small enough $\alpha$  even if the parameters has been shifted to those
in the transmitter, i.e.  $(\sigma, r, b)=(16.0, 45.6, 4.0)$. 
An example of the  synchronization and parameter recovery process is shown in Fig. 11 for 
$(\epsilon, \alpha)=(100, 0.2)$. The convergence rate  with the additional parameter adaptive 
control  is  slower than that  without  these control loops.  Now if the bit duration in the transmitter
is longer than the synchronization transient,   the attacking system can follow the parameter 
modulation in the transmitter and thus decode the message. An example is shown in Fig. 12, where
the bit duration is 16 ms. While the parameter $b$ clearly follows the modulation, the other two parameters
also reflect the switch of the message from one value to the other.  
It is seen again in Fig. 11 that the synchronization error decreases exponentially, with fluctuations only
within very small time scale, so that the  message decoding is  also very robust to channel noise. 
Fig. 13 shows the recoved parameters  when the transmitted signal $s=x_1$ contains an additive noise
in $[-1, 1]$.

\bigskip
\section{Discussion}

Based on the assumption that the chaotic system structure is in the public domain and the 
system parameters are kept in secret as  the encryption key in a secure 
communication system encoding digital message by parameter modulation,  
we have shown  how an  intruder might  decode the message  
using  an appropriate attacking system  with parameter adaptive control by the 
transmitted signal, even though the intruder does not have access to the exact parameter
values in the transmitter. A requirement  for the success of this attacking approach is that  
the intruder can design a synchronizing parameter adaptive control system which is 
quite robust to  mismatches between the  parameters of the two systems, so that the message
can be recoved without resorting to the exact parameters in the transmitter, but within some 
neighborhood. Based on the knowledge of the system, it is practically possible for the intruder
to get into such a neighborhood using some system identification methods.  

For some systems, such a robust synchronizing intruder system with parameter  adaptive control 
can be constructed based on an  analysis of a global Laypunov function. Generally, such
an analytical treatment is impossible, and we proposed a quite general and practical local 
Lyapunov function method 
to design parameter adaptive control rules based on a system which has been synchronized by the
transmitted signal. Such a synchronizing system is often obtainable using many possible approaches
for constructing synchronization chaotic systems, such as Pecora-Carroll decomposition, active-passive
 decomposition
or feed-back control. In many systems, the synchronization is robust to parameter 
mismatches  if the coupling is not close to the synchronization threshold. 
The parameter control rules are designed by seeking  appropriate functions of the
synchronization error whose time average change smoothly when the parameters in the originally synchronized
systems  deviate slightly from each other. The smooth control surface is obtained in simulation or
experiment by perturbing the parameters that will be involved in adaptive control. Although this scheme
is quite general, in practice, finding a set of appropriate functions may not be a trivial task 
 when many parameters are involved in modulation. In some cases,  the originally synchronized 
system may be very sensitive to parameter mismatches due to unstable invariant sets embedded in the
synchronization manifold~\cite{hcp, gb}, and the proposed local Lyapunov function may not be successful in 
designing additional parameter adaptive control loops for such systems. However, such systems will not be
used in the communication systems because the authorized receiver cannot decode the message correctly
in practical environment with unavoidable perturbations.

Employing  some system identification methods, 
the intruder may be able to identify  a region near the transmitter parameters 
in the parameter space in order to design a stable intruder system. Furthermore, 
the intruder may be able to get close enough to the transmitter parameters 
by monitoring  the synchronization error  while scanning the parameter space,
so that the message can be decoded with very low rate of errors.
During the decoding process, 
the  intruder can improve the decoding by comparison of the results using different
parameters in the identified  region.  
In some systems,  it is also possible to design adaptive control loops for all the 
system parameters, so that the message can be decoded  even more reliablely.

Based on our investigation, we would like to point out an interesting paradox between 
robustness and security in chaotic communications. 
Since in practice, parameter mismatches  and external noise is unavoidable, we would
require the synchronization systems to be robust to these perturbations, so that 
high-quality synchronization can be established between the transmitter and 
the authorized 
receiver to recover the message correctly in practical implementation. 
On the other hand, this robustness may be 
employed by a third party to compromise  the security. How to improve the security while 
maintaining the robustness is an interesting and  meaningful research topic for chaotic 
communications.

\bigskip
{\bf Acknowledgements:}

This work was supported in part by research grant No. RP960689 at the National
University of Singapore.  Zhou is supported by NSTB.

\newpage

Table I. Values of the parameters of the Chua's circuit in the transmitter.\\

\begin{tabular}{cccccccc}\hline
$C_1$ (nF) & $C_2$ (nF) & $R$ ($\Omega$) & $L$ (mH) & $G_0$ (ms) & $G_1$ (ms) & $1/r$ ($\mu$s) &
 $B_p$ (V)\\ \hline
10       &  100       & 1680           & 18       &  $-$0.753    &  $-$0.396    &  6
  & 1   \\ \hline\\
\end{tabular}

\bigskip

Table II. Parameters of the transmitter and the intruder and their relative differences.\\

\begin{tabular}{c|ccccc}\hline
transmitter      & $C_1$        & $C_2$       & $R$        & $L$       & $B_p$\\ \hline
intruder         & $0.817C_1$   & $1.163C_2$  & $1.072R$   & $0.897L$  & $0.849B_p$\\ \hline
differences ($\%$)&  $-$18.3         & 16.3       &  7.2       & $-$10.3    & 15.1     \\\hline
\end{tabular}

\newpage 
\vspace{1cm}
{\bf Figure Captions}

\begin{description}

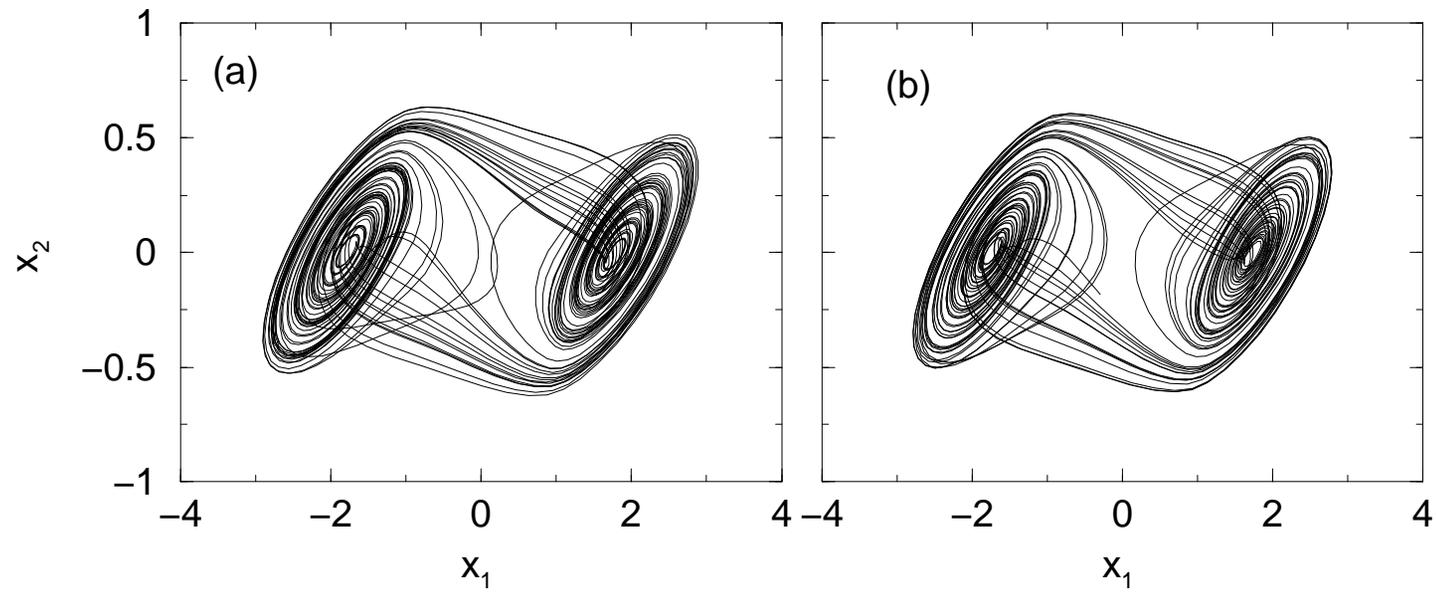
\item Fig. 1.   Two attractors used to encode binary information, with (a)
corresponding to bit $+1$ and (b) to $-1$.

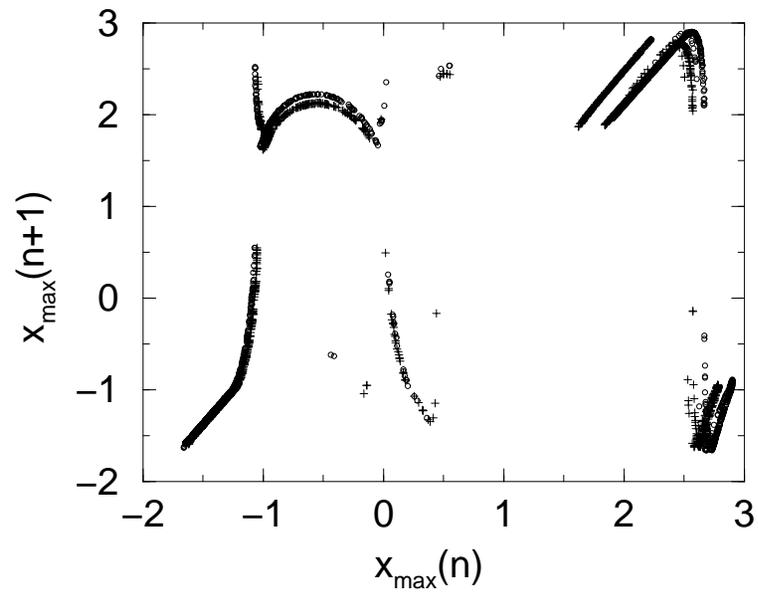
\item Fig. 2. Return maps of the consecutive maxima of the transmitted signal $x_1$.
        Most of the points for the attractor in Fig. 2(a) (circles) 
        and those for the attractor in Fig. 2(b) (crosses) do not 
        have distinguishable separation. 

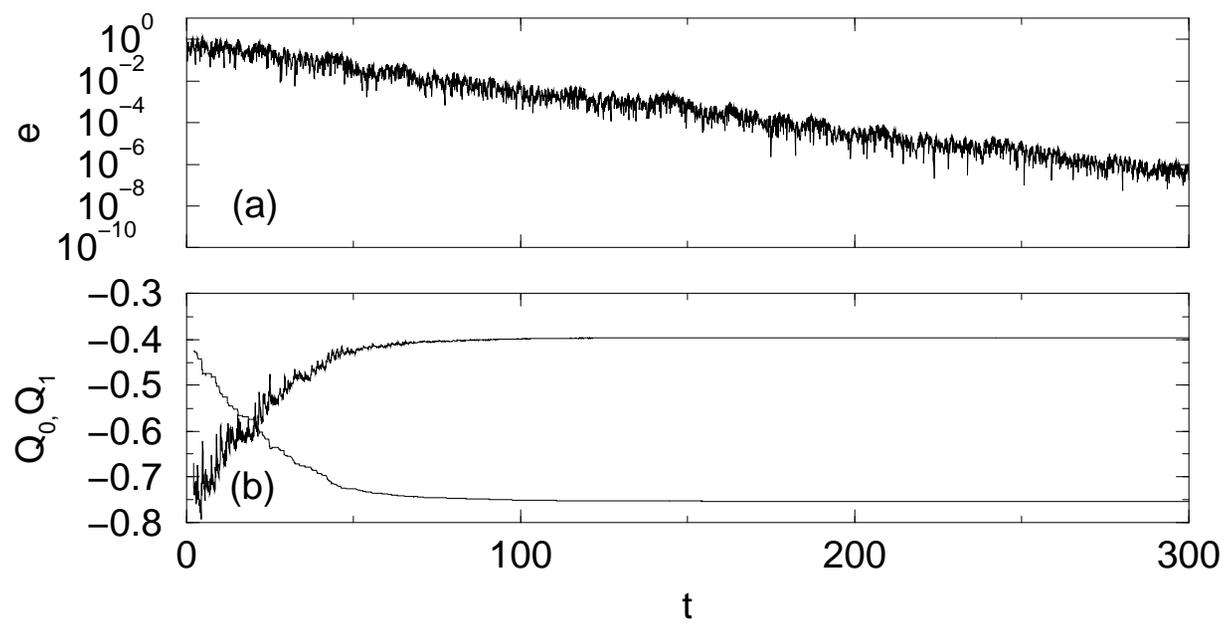
\item Fig. 3. Synchronization process of the  intruder  to the attractor in Fig. 2(a)

       (a) Synchronization error $\sqrt{\sum e_i^2}$. 
       (b) Convergence of parameters $Q_0$ and $Q_1$ to those of the transmitter 
           $G_0=-0.753$ and  $G_1=-0.396$, respectively.
        In all the figures in this paper, the unit of time is ms.

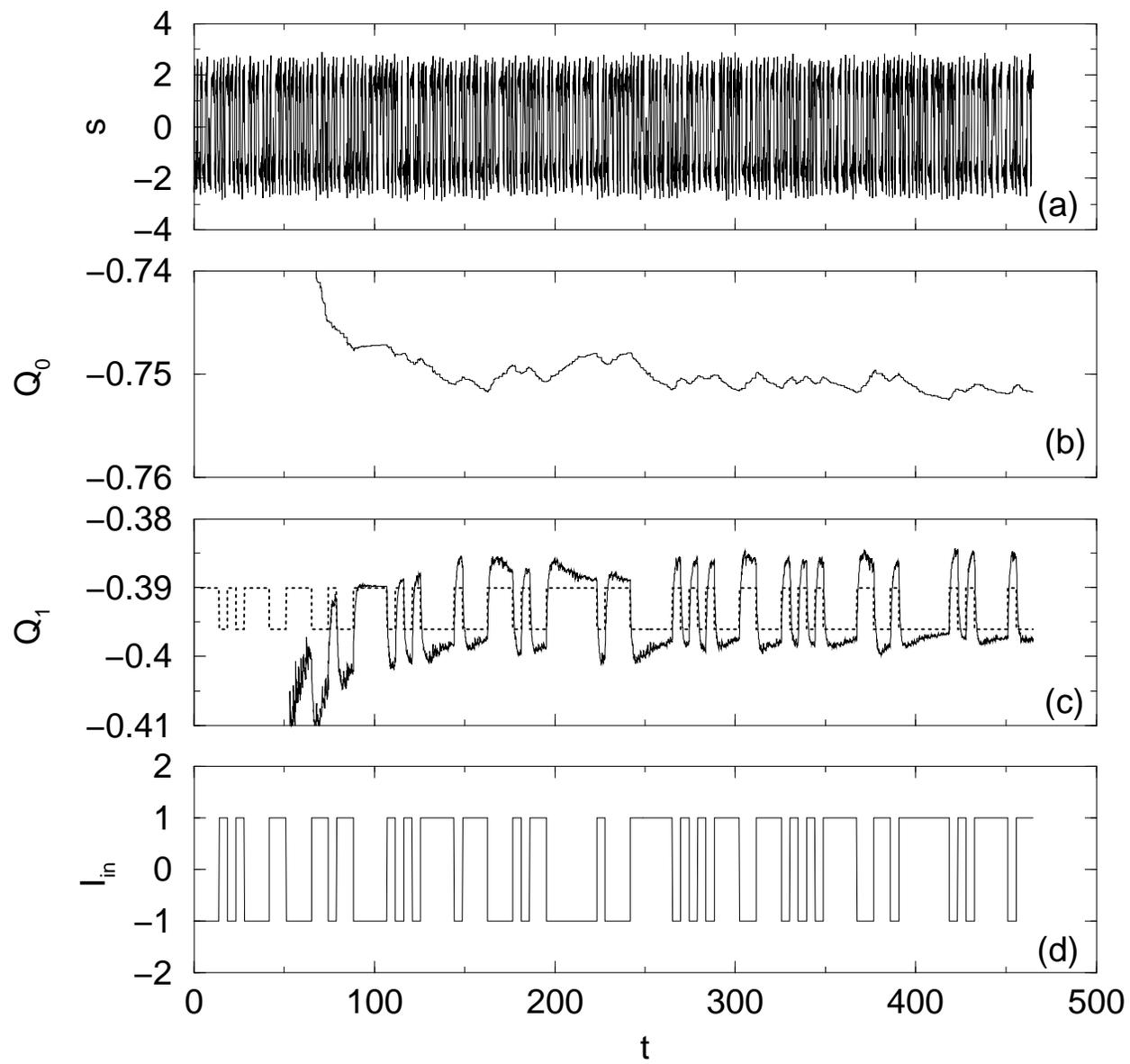
\item Fig. 4. The process of following the parameter modulation in the transmitter. 

        (a) The transmitted signal $s=x_1$.
        (b) Change of $Q_0$.
        (c) Change of $Q_1$. The dotted line shows the parameter modulation in the
transmitter.

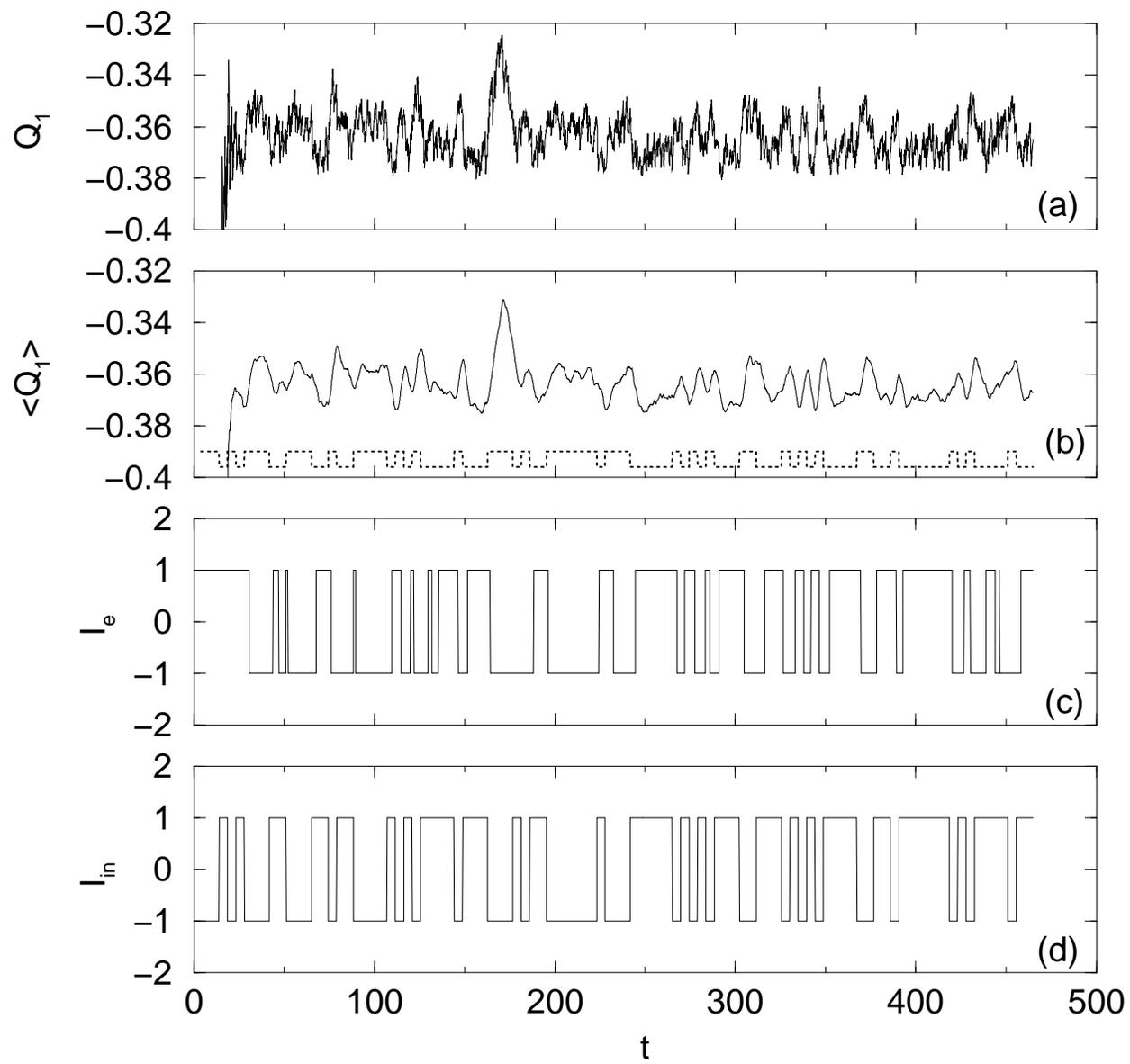
\item Fig. 5.  An example of decoding process.
         
        (a) Evolution of $Q_1$. It oscillates noisely due to quite large  parameter 
            mismatches.
        (b) Smoothed $Q_1$ by moving average with length of 4 ms.
            The dotted line shows the parameter modulation in the transmitter.
        (c) The decoded message by threshold testing with  a threshold $Q_{th}=-0.365$.     
        (d)  The original message stream.        

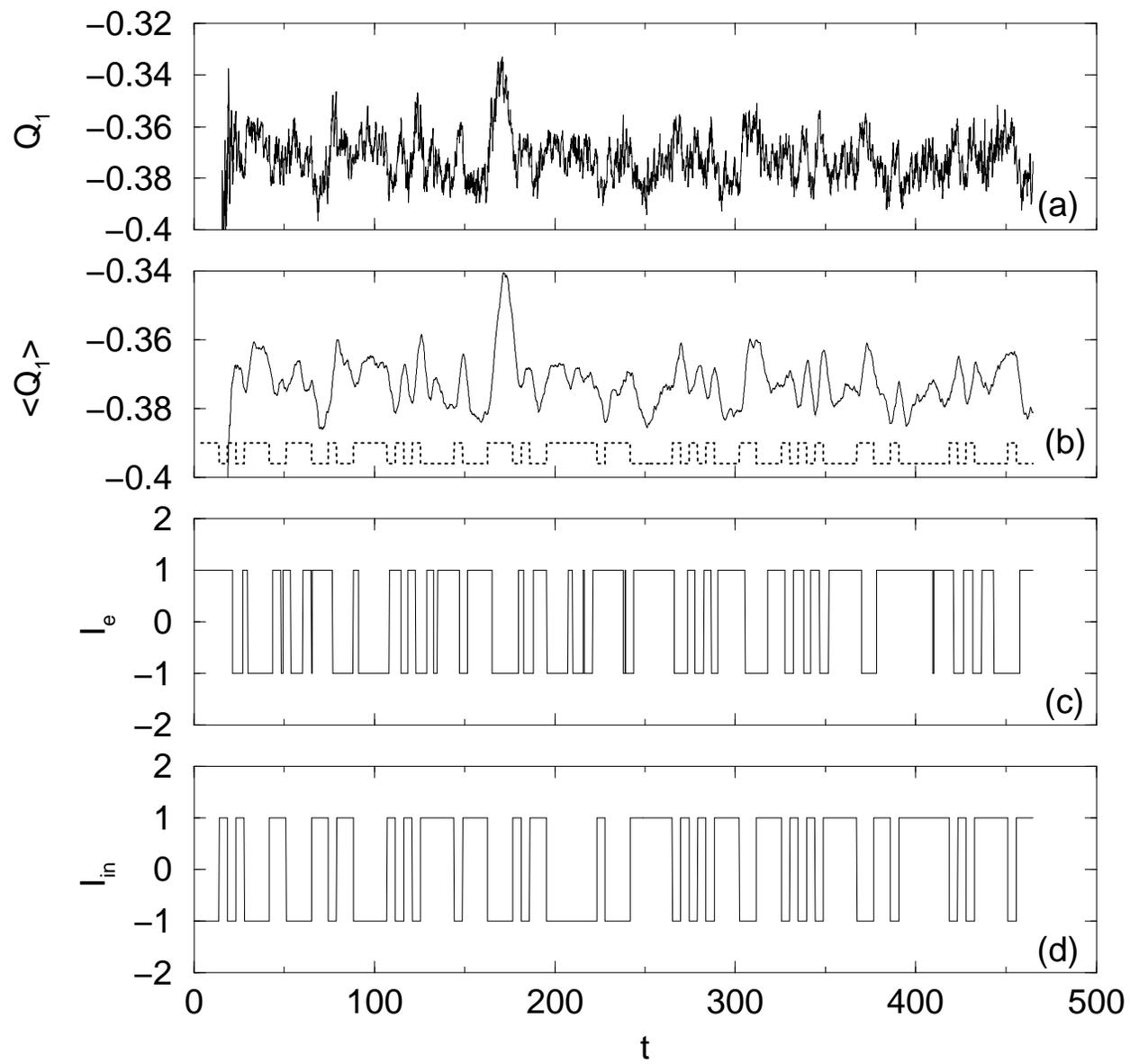
\item Fig. 6.  Analogous to Fig. 6, but with $s$ containing  noise 
           between $[-0.2,0.2]$.

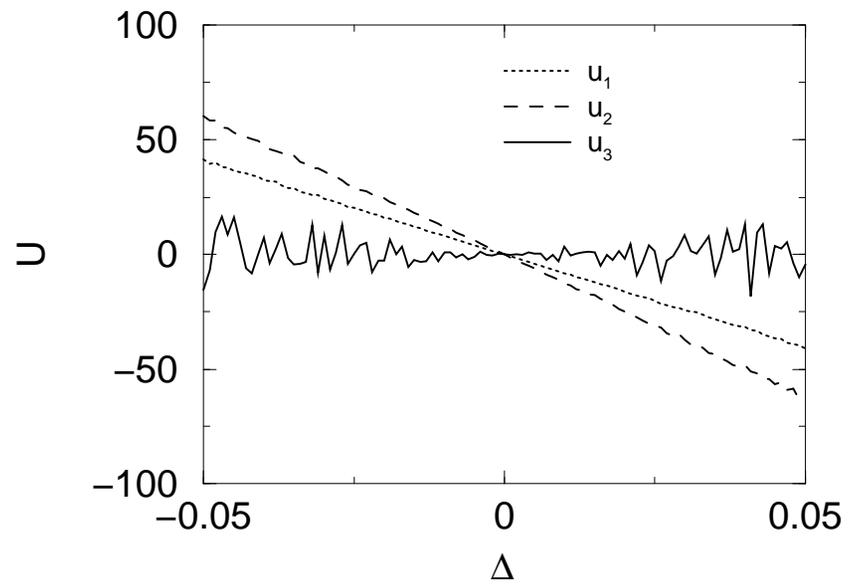
\item Fig. 7. $U$ as a function of the relative deviation of the parameter $b$ for different
             choice of function $u$. The smooth functions can be used to design parameter adaptive
             control loop.

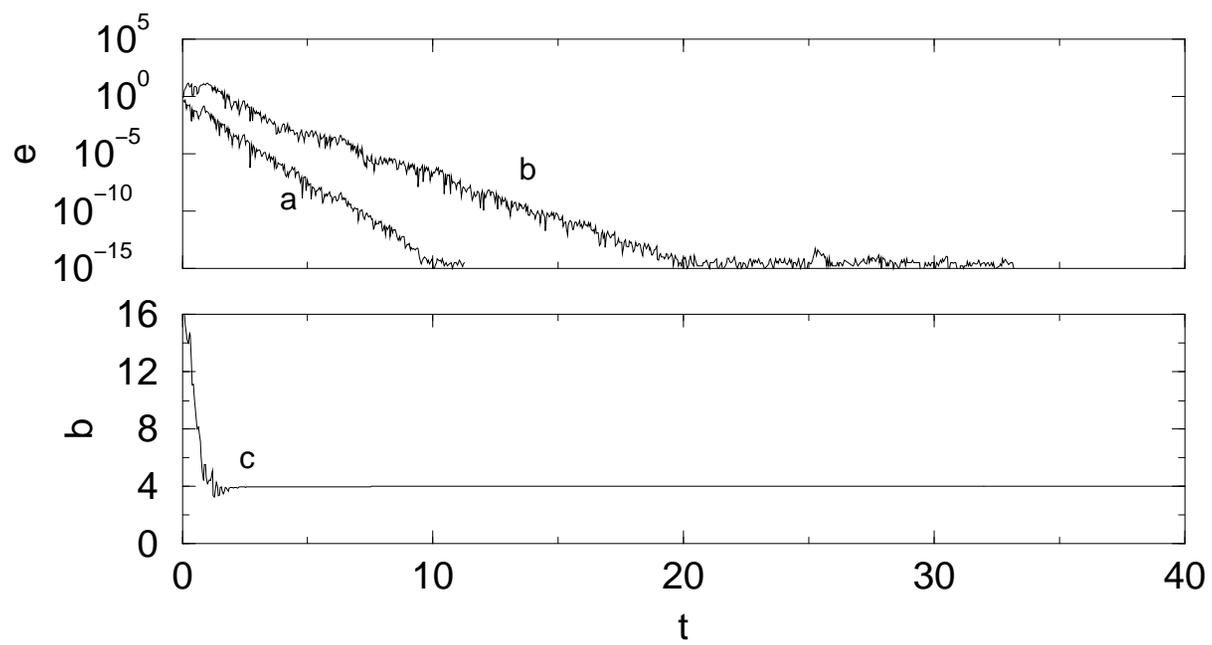
\item Fig. 8. Synchronization and parameter recovery with the additional parameter adaptive control
           loop of Eq. (35).
                  (a) and (b) are  synchronization errors without  and with this control loop respectively.
                  (c) is the convergence of the parameter $b$.

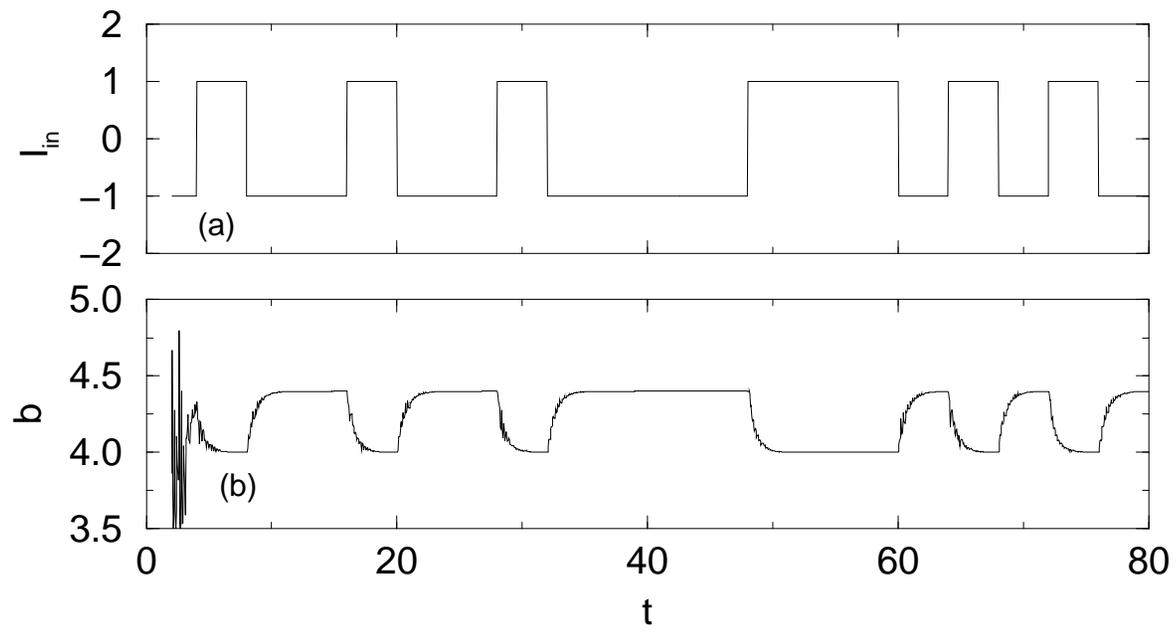
\item Fig. 9. The process of following the parameter modulation in the case that the rest of the
              parameters $\sigma$ and $r$ are identical. 

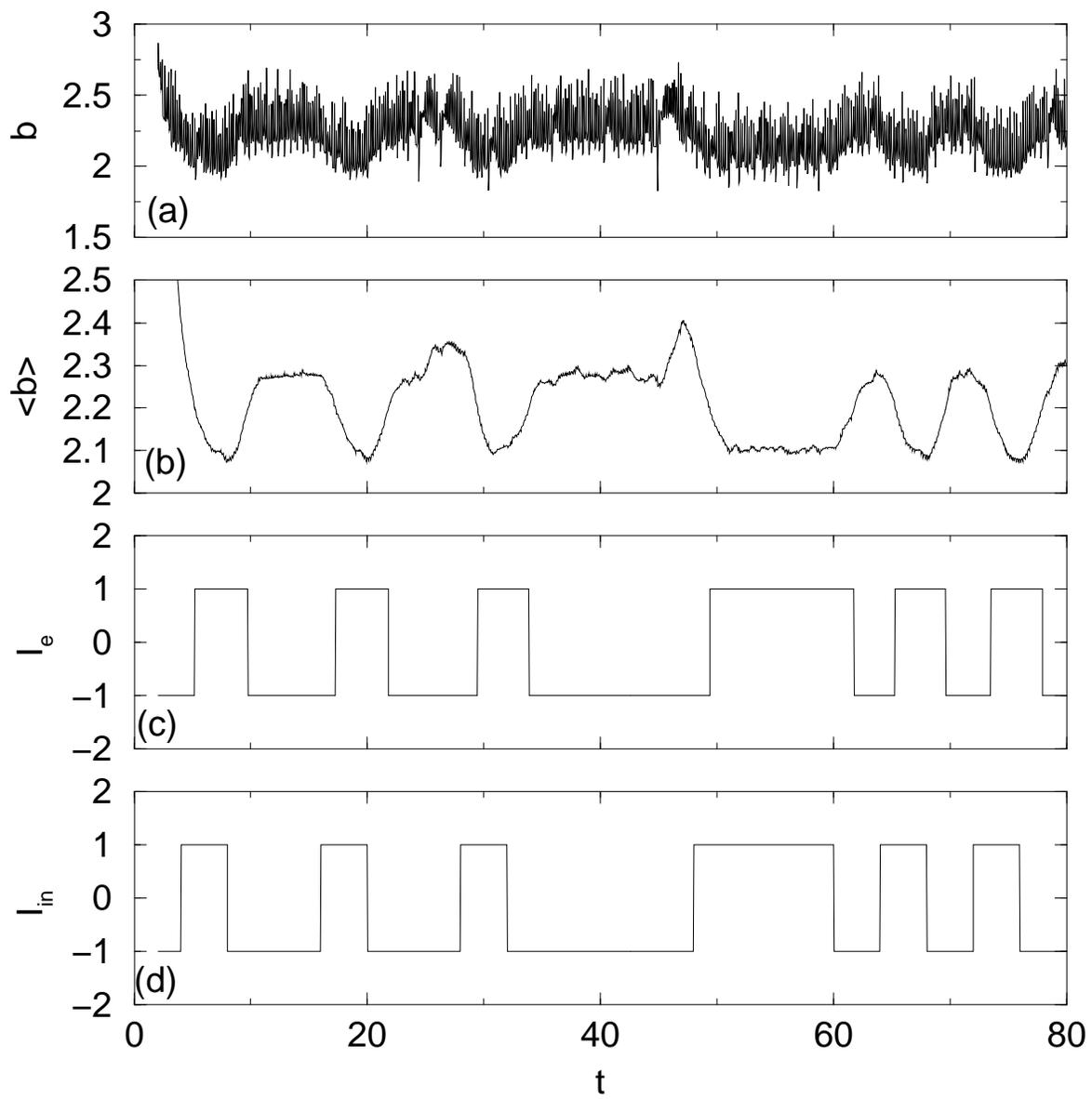
\item Fig. 10. Illustration of message decoding when the parameters  $\sigma$ and $r$  have large 
               mismatches between the systems. 
               (a) Evolution of $b$.
               (b) smoothed $b$ by moving average filter  with length of 2ms.
               (c) The decoded message by threshold testing with  a threshold $b_{th}=2.2$. 
               (d)  The original message stream.

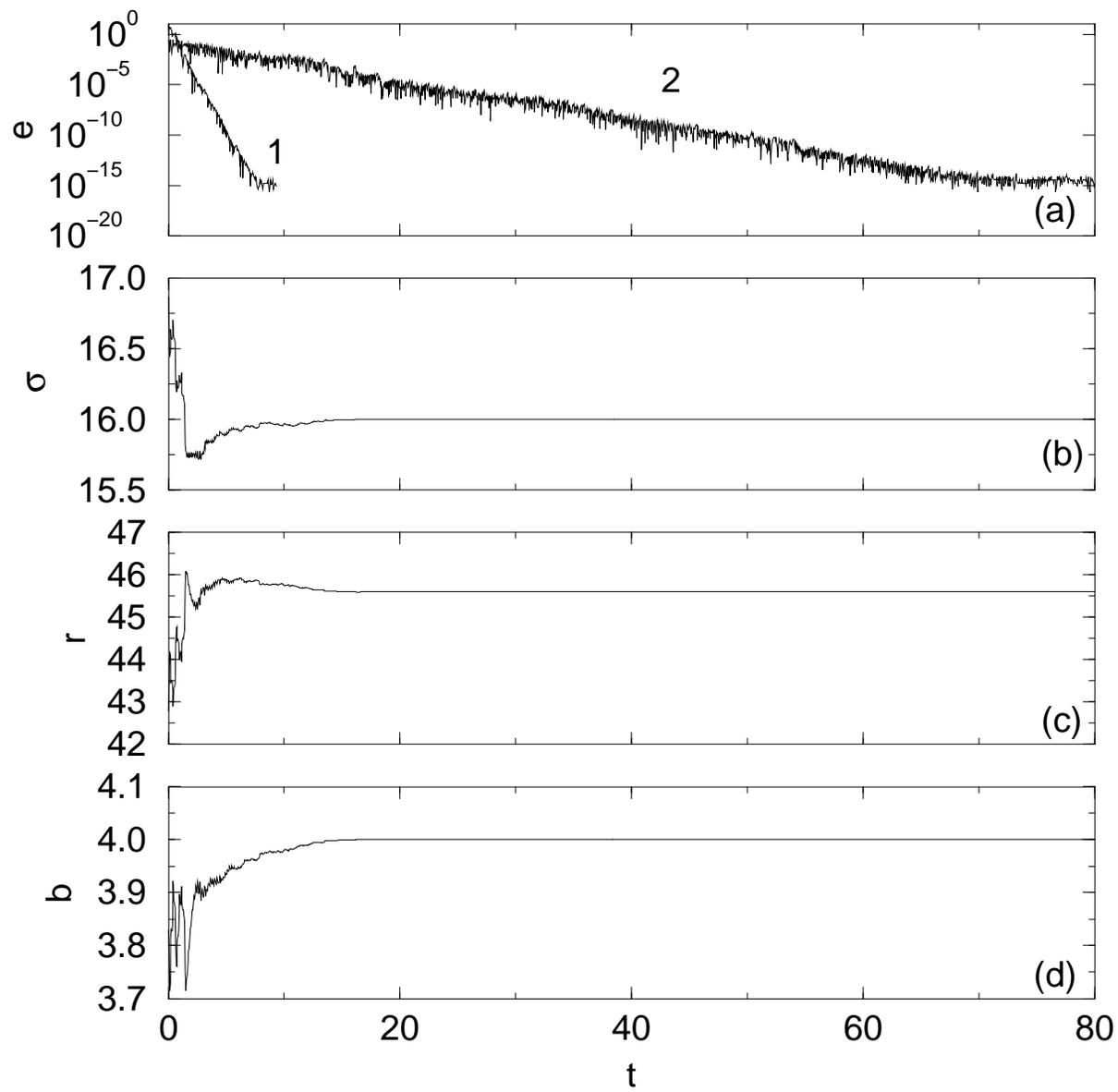
\item Fig. 11. Synchronization and parameter recovery with the additional parameter adaptive control
               loops for all the three parameters.  
               (a) Plots  1 and 2 are  synchronization errors without  and with these control loops respectively.
                (b), (c) and (d) are the evolutions of the parameters $\sigma$, $r$ and $b$, respectively. 
           
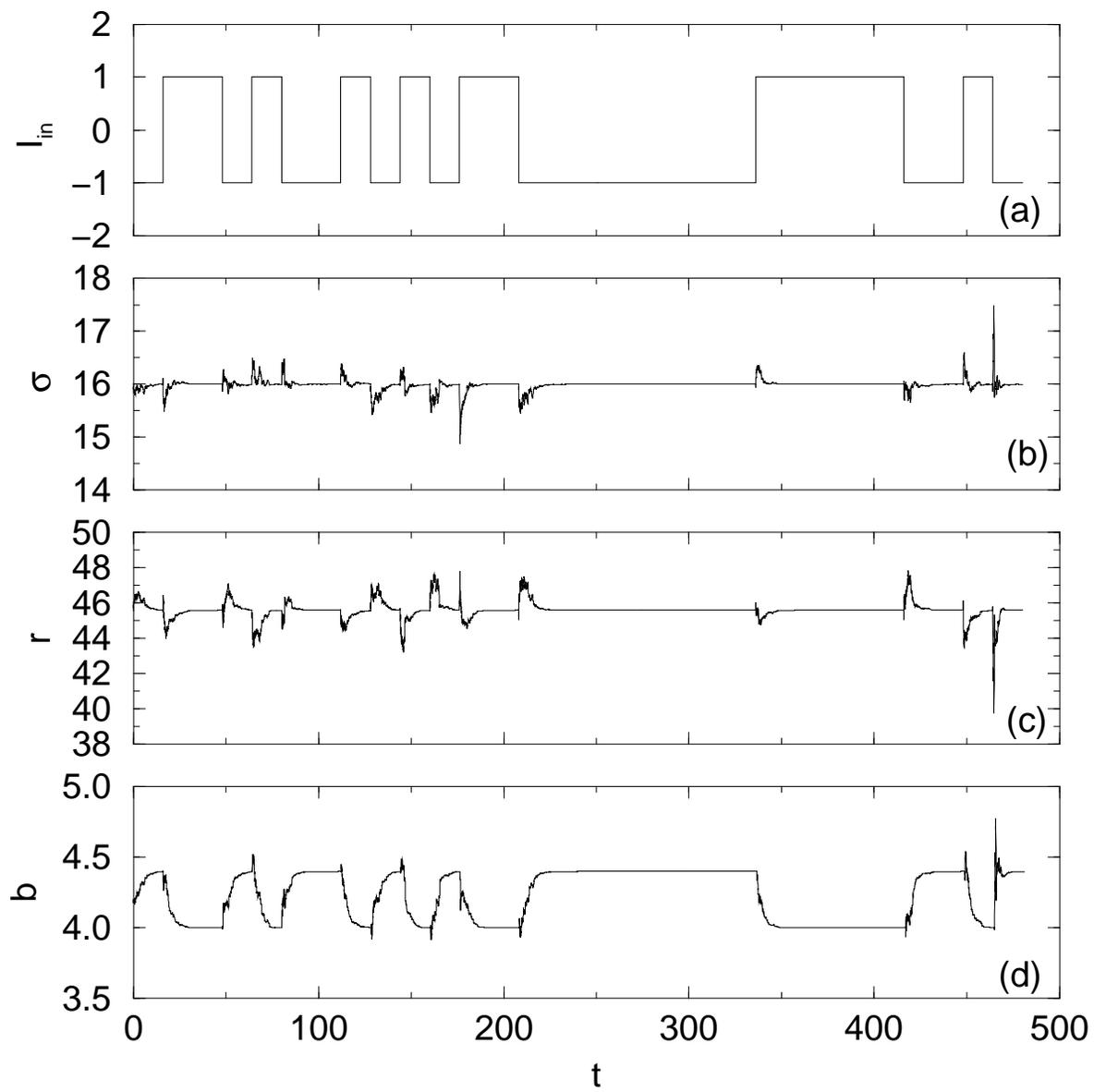
\item Fig. 12. The process of following the parameter modulation in the transmitter. 
               (a) is an input message stream.
               (b), (c) and (d) are the evolutions of the parameters $\sigma$, $r$ and $b$, respectively.
               The bit duration is 16 ms.
 
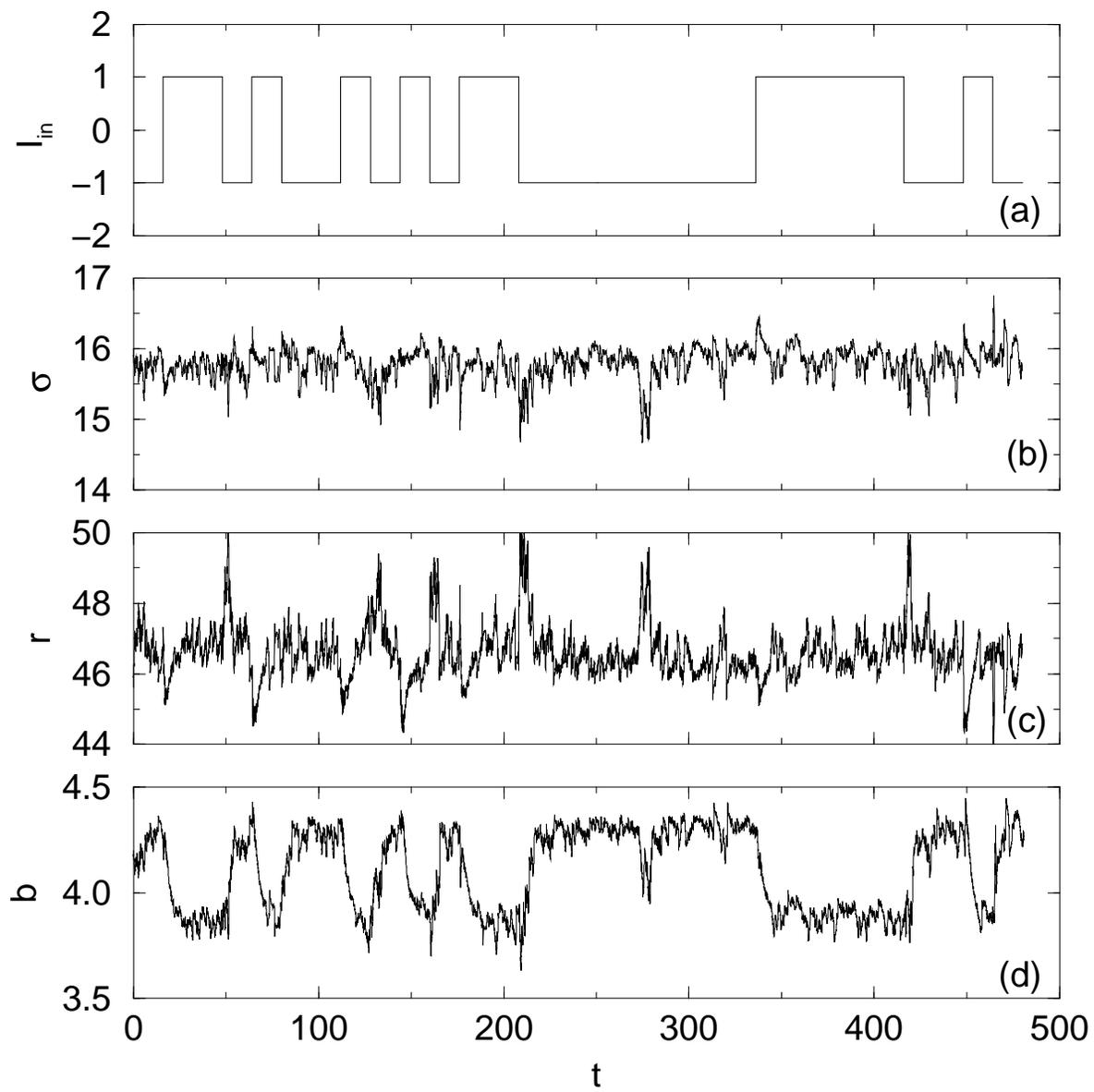
\item Fig. 13. Robustness of the message decoding in the presence of channel noise within $[-1, 1]$.

\end{description}
\end{document}